\documentclass[12pt]{iopart}
\usepackage{enumerate}
\usepackage{cite}

\usepackage{color}
\usepackage{graphicx}
\usepackage{amssymb}
\usepackage{bm}
\usepackage{epsfig}

\begin{document}

\title[Annihilation vs.~Decay]{Annihilation vs.~Decay: Constraining dark
  matter properties from a gamma-ray detection}

\author{Sergio Palomares-Ruiz}
\address{Centro de F\'{\i}sica Te\'orica de Part\'{\i}culas, Instituto
  Superior T\'ecnico, \\
Av. Rovisco Pais 1, 1049-001 Lisboa, Portugal} 
\ead{sergio.palomares.ruiz@ist.utl.pt}

\author{Jennifer M.~Siegal-Gaskins}
\address{Center for Cosmology and Astro-Particle Physics, The Ohio State
  University, \\
191 W. Woodruff Ave., Columbus, OH 43210, USA} 
\ead{jsg@mps.ohio-state.edu}

\begin{abstract}
Most proposed dark matter candidates are stable and are produced
thermally in the early Universe.  However, there is also the
possibility of unstable (but long-lived) dark matter, produced
thermally or otherwise.  We propose a strategy to distinguish between
dark matter annihilation and/or decay in the case that a clear signal
is detected in gamma-ray observations of Milky Way dwarf spheroidal 
galaxies with gamma-ray experiments.  The sole measurement of the
energy spectrum of an indirect signal would render the discrimination
between these cases impossible.  We show that by examining the
dependence of the intensity and energy spectrum on the angular
distribution of the emission, the origin could be identified as decay,
annihilation, or both.  In addition, once the type of signal is
established, we show how these measurements could help to extract
information about the dark matter properties, including mass,
annihilation cross section, lifetime, dominant annihilation and decay 
channels, and the presence of substructure.  Although an application
of the approach presented here would likely be feasible with current
experiments only for very optimistic dark matter scenarios, the improved
sensitivity of upcoming experiments could enable this technique to be
used to study a wider range of dark matter models.

\vspace{3ex}
\noindent
{\bf Keywords:} dark matter theory, gamma-ray theory, dwarf galaxies
\end{abstract}

\pacs{95.35.+d, 95.85.Pw, 98.56.Wm, 98.62.Gq
\hfill CFTP/10-003}

\submitto{Journal of Cosmology and Astroparticle Physics}

\maketitle

\section{Introduction}
\label{introduction}

Many different astrophysical and cosmological observations have found
evidence for the existence of non-luminous, non-baryonic dark matter
(for reviews see, e.g., Refs.~\cite{Jungman:1995df, Bergstrom:2000pn,
  Bertone:2004pz}) and indicate that it constitutes about 80\% of the
mass content of the Universe~\cite{Dunkley:2008ie}.  Despite the
precision with which the cosmological dark matter density has been
measured, little is known about the origin and properties of the dark
matter particle, such as its mass, spin, couplings, and its
distribution on small scales.    

Several candidate dark matter particles have been proposed with masses
from the electroweak scale to superheavy candidates at the Planck scale
(see, e.g., Refs.~\cite{Bertone:2004pz, Bergstrom:2009ib} for a
comprehensive list).  Light particles have also been considered as
possible dark matter candidates: axions~\cite{axions}, sterile
neutrinos with masses in the keV range~\cite{Dodelson:1993je} and
light scalars with MeV--GeV masses~\cite{LDM, Boehm:2003bt, MeVDM,
  Boehm:2006mi}.  However, the most popular candidates are weakly
interacting massive particles (WIMPs) which arise in extensions of the
Standard Model such as supersymmetric (e.g.,
Ref.~\cite{Jungman:1995df}), little Higgs (e.g.,
Ref.~\cite{Birkedal:2006fz}), or extra-dimensions models (e.g.,
Ref.~\cite{Hooper:2007qk}).

Most proposed WIMPs are stable and are produced thermally in the early
Universe with an annihilation cross section (times relative velocity)
of $\langle\sigma v\rangle \sim 3 \times 10^{-26}$~cm$^3$~s$^{-1}$.
However, dark matter may be unstable but long-lived; the only
requirement in order for it to be a thermal relic and present today is
that it has a lifetime $\tau_\chi$ much longer than the age of the
Universe $t_{\rm U} \simeq 4 \times 10^{17}$~s.  It is also possible
that dark matter is not a thermal relic, which would allow it to have
a larger annihilation cross section than the canonical value for WIMP
thermal relics.

Although the case of a non-thermal~\cite{Kaplinghat:2000vt, unitarity}
or unstable~\cite{decaymodels} dark matter candidate was considered
several decades ago, recently a great deal of interest in these
scenarios has been generated by the rise in the positron fraction in
the tens of GeV range measured by the PAMELA experiment~\cite{PAMELA}.
One possibility to explain the PAMELA data is by the injection of
positrons by annihilation~\cite{Cirelli:2008pk, Barger:2008su,
  PAMELAannihilationexotic, Grasso:2009ma}, decay~\cite{PAMELAdecay,
  Nardi:2008ix} or both~\cite{Cheung:2009si} of dark matter in the
solar neighborhood (see also Ref.~\cite{He:2009ra} and references
therein).  In the case of annihilation in the smooth halo,
enhancements to the annihilation cross-section of the order of
$10$-$10^5$ are required (with respect to thermal dark matter with no
Sommerfeld or Breit-Wigner enhancements and assuming there is no
nearby dark matter clump)~\cite{Cirelli:2008pk}.  However, note that the
contribution from substructure could be significant~\cite{Brun:2009aj}
and interestingly, the cumulative effect from distant subhalos could
modify the electron and positron spectra at our galactic
position~\cite{Cline:2010ag}.

Indirect dark matter searches look for the products of dark matter
annihilation or decay, which include not only antimatter, as in the
case of PAMELA, but also neutrinos~\cite{Rott:2009hr, indirectnusun}
and photons.  In particular, targets for indirect searches in
gamma-rays include dark matter in extragalactic
structures~\cite{cosmoindirect}, the Galactic
Center~\cite{GC, Albert:2005kh, Aharonian:2006wh, Baltz:2008wd}, the
Milky Way halo~\cite{Baltz:2008wd, MWhalo, Pieri:2009je, Abdo:2010nc},
its subhalos~\cite{Baltz:2008wd, Pieri:2009je, subhalos,
  Kuhlen:2008aw} and known dwarf galaxies~\cite{Baltz:2008wd, dG,
  Evans:2003sc, Strigari:2007at, Essig:2009jx, MAGICdG, HESSdG,
  Driscoll:2008zz, Wagner:2009wp, Abdo:2010ex}.

Different approaches have been proposed to constrain dark matter
properties by using indirect or direct measurements or their
combination~\cite{direct, indirectnuprop, indirectgamma,
  Hensley:2009gh, Beltran:2008xg}.  To extract the properties of the
dark matter particle from the detection of an indirect signal requires
several pieces of information.  While the energy spectrum of the
signal is dependent on the dark matter properties (mass $m_{\chi}$,
annihilation cross section $\langle\sigma v\rangle$, lifetime
$\tau_{\chi}$, and annihilation and/or decay channels), sufficient
degeneracies exist to prevent accurate reconstruction of the dark
matter properties from the energy spectrum alone.  In particular, the
sole measurement of the energy spectrum would make it impossible to
know if the indirect signal from dark matter is produced by
annihilation or decay.  The spectrum of the former is characterized by
a cutoff at an energy equal to the dark matter mass, while the cutoff
in the spectrum from the latter is at an energy equal to half of the
dark matter mass.

In this work we address the following two questions: (1) in the case
that an indirect dark matter signal is detected in gamma rays, can
annihilation and/or decay be identified as the origin of the signal?
and (2) what information about the particle properties can be obtained
from the indirect measurement? If dark matter is unstable and produces
an observable signal from decay, an annihilation signal may also be
present.  In principle, there is a range of parameters for which the
two signals would be comparable, which would present a challenge for
distinguishing the cases of annihilation, decay, and those where both
annihilation and decay are significant contributors to the measured
signal.

In order to tackle this problem, we propose a strategy to distinguish
between these scenarios using gamma-ray observations of Milky Way dwarf
spheroidal galaxies with current or future gamma-ray telescopes.
Current missions include the Large Area Telescope (LAT) aboard the
Fermi Gamma-ray Space Telescope (\emph{Fermi})~\cite{Atwood:2009ez},
which observes gamma-rays in the range from 20~MeV to greater than
300~GeV, and atmospheric \v{C}erenkov telescopes (ACTs) such as
HESS~\cite{Hinton:2004eu}, VERITAS~\cite{Holder:2008ux}, and
MAGIC~\cite{Lorenz:2004ah}, which observe emission above
$\sim$100~GeV, and the planned ground-based ACT arrays
CTA~\cite{Wagner:2009cs} and AGIS~\cite{Buckley:2009zza}.  We
demonstrate that, in the case that a gamma-ray signal is clearly
detected, the origin could be identified as decay, annihilation, or
both by examining the dependence of the intensity and energy spectrum
on the angular distribution of the emission.  Furthermore, if
annihilation and decay each contribute significantly to the signal, we
show how these observations could be used to extract information about
the dark matter mass, lifetime, annihilation cross section, and
dominant annihilation and decay channels.  In addition, as a byproduct
of this analysis, one might also establish or limit the contribution
to the signal from substructure in the dwarf galaxy's halo. 

The paper is organized as follows.  We outline our proposed observing
strategy in \S\ref{method}.  In \S\ref{dwarfs} we summarize the
properties of the dwarf galaxies we consider and describe our approach
to modeling their dark matter halos and subhalos.  The gamma-ray
spectra from dark matter annihilation and decay are discussed in
\S\ref{spectra}.  In \S\ref{results} we demonstrate the proposed
method by presenting example results for various dark matter scenarios
for selected dwarf galaxies.  We summarize our results and conclude in  
\S\ref{conclusions}.

\section{Method}
\label{method}

An indirect signal from annihilation or decay originates from the same
dark matter particles, but these two processes give rise to different
angular distributions of the emission and different energy spectra.
We propose an observing strategy to distinguish these two processes by
the angular variation of the intensity and the energy spectrum of the
signal.  We first describe the general case of gamma-ray observations
of an external halo (specifically, we assume that the distance from
our position to the object is large compared to the size of the region
in the object from which the signal originates), and then illustrate
the technique for specific dwarf galaxies. 

As pointed out in Refs.~\cite{Bertone:2007aw, PalomaresRuiz:2007ry}, 
angular information is crucial to distinguish dark matter annihilation
from decay.  The rate of annihilation scales as the square of the dark
matter density $\rho$, while that of decay scales linearly with the
density, and consequently the angular distribution of the signal from
dark matter annihilation in an external halo is expected to follow a
steeper profile than that from dark matter decay.  However the spatial
distribution of dark matter substructure in a halo also scales roughly
as $\rho$ in the outer regions of the halo.  
Consequently, annihilation in this component could produce a
similar flattening in the angular distribution of the observed
emission as is expected for decay (see, e.g.,
Fig.~\ref{fig:iprofiles}, discussed in \S\ref{dwarfs}).

From an observational standpoint, a dramatic decrease in the observed
intensity between the center of the object and that at larger angles
is a clear indicator of the simple case of annihilation in the smooth
halo only, while the observation of a shallow emission profile at all
angles would strongly suggest decay only.  On the other hand, the
observation of a bright central region but with the intensity falling
off more slowly in the outer regions is less straightforward to
interpret, as it could indicate annihilation with an important
contribution from substructure, or both annihilation and decay
contributing significantly.  In this case, we demonstrate that an
analysis of the energy spectrum of the signal as a function of angular
distance from the center of the object could provide the necessary
information to distinguish these possibilities. 

If only one process (annihilation or decay) produces a detectable
signal, the energy spectrum of the dark matter signal is the same from
all regions of the object, with the intensity varying according to how
the rate of that process depends on the dark matter distribution.  If
both processes produce detectable signals, the energy spectrum of the
total signal varies according to the contribution from each process.
With generality, we can assume that in this two-process scenario the
annihilation signal is always dominant in the inner regions of the
object, with decay becoming more important at larger angles from the
center of the object.  Thus, we identify that both annihilation and
decay are present by observing a change in the energy spectrum of the
signal as a function of angle.  In the following we further examine
the information available from an indirect measurement in the
two-process case. 

The differential intensity of the gamma-ray signal (photons per time
per area per solid angle per energy) at an angle $\psi$ from the
center of the object from dark matter decay or annihilation can be
written as the product of a term depending on the particle properties
$P$ and a term depending on the dark matter distribution $\Phi$, 
\begin{equation}
\label{eq:intenscompact}
I_{x}(\psi)=P_{x} \times \Phi_{x}(\psi)
\end{equation}
where $x=$ D or A, for decay and annihilation, respectively.  The
particle physics factors are defined as 
\begin{eqnarray}
\label{eq:pfactors}
P_{\rm D}& = & \frac{1}{m_{\chi}\tau_{\chi}} \frac{{\rm d}N_{\rm
    D}}{{\rm d}E}\\  
P_{\rm A} & = & \frac{\langle \sigma v\rangle}{2m_{\chi}^{2}}
\frac{{\rm d}N_{\rm A}}{{\rm d}E}~, 
\end{eqnarray}
where $m_{\chi}$ is the mass of the particle, $\tau_{\chi}$ is the
lifetime, $\langle \sigma v\rangle$ is the annihilation cross section,
and ${\rm d}N_{\rm D}/{\rm d}E$ and ${\rm d}N_{\rm A}/{\rm d}E$ are
the differential photon spectra per decay or annihilation,
respectively.  The $\Phi_{x}(\psi)$ are determined by the dark matter
density profile which can be estimated from kinematic data, and the
${\rm d}N_{x}/{\rm d}E$ are set by the channels for each process.
However, we emphasize that in general this is insufficient information
to determine the particle physics properties, since for a given final
state, the mass of the dark matter particle that produces the observed
gamma-ray spectrum is a factor of 2 larger in the case of decay than
in annihilation.  The angular dependence of the spectral information
is essential to distinguish the two processes and break the
degeneracy. 

Since there are parameters that determine the intensity from
annihilation and decay that are common to both processes, we can
distill the information contained in the indirect signal by
considering the ratio of the intensity at a given $\psi$ from 
annihilation to decay, 
\begin{equation}
\label{eq:ratio}
\frac{I_{\rm D}}{I_{\rm A}}(\psi) = \frac{P_{\rm D}}{P_{\rm A}} \,
\frac{\Phi_{\rm D}(\psi)}{\Phi_{\rm A}(\psi)} = 
\frac{2m_{\chi}}{\langle \sigma v\rangle \tau_\chi} \, 
\frac{({\rm d}N_{\rm D}/{\rm d}E)}{({\rm d}N_{\rm A}/{\rm d}E)} \,
\frac{\Phi_{\rm D}(\psi)}{\Phi_{\rm A}(\psi)}~. 
\end{equation}
For a specified density profile and set of annihilation and decay
channels the ratio of the intensities of the two signals at a given
angle depends only on the particle properties.  Defining the angle at
which the two signals are equal at a given energy (or, in practice,
integrated over some energy range) as $\psi_{\rm cross}$, we can write 
\begin{equation}
\label{eq:taucross}
\tau_\chi = \frac{2m_{\chi}}{\langle \sigma v\rangle}\, \frac{\int dE \,
  ({\rm d}N_{\rm D}/{\rm d}E)}{\int dE \, ({\rm d}N_{\rm A}/{\rm d}E)}\,
\frac{\Phi_{\rm D}(\psi_{\rm cross})}{\Phi_{\rm A}(\psi_{\rm cross})}~. 
\end{equation}
A measurement of $\psi_{\rm cross}$ thus determines the value of the
lifetime in terms of the mass and annihilation cross section.
However, a key point in this scenario is that by measuring $\psi_{\rm
  cross}$, the presence of both annihilation and decay is confirmed,
so by examination of the signal in the inner and
outer regions of the object, the degeneracy in the dark matter particle
mass $m_{\chi}$ could be broken.  In this case the particle properties
$\tau_\chi$ and $\langle \sigma v\rangle$ are also uniquely determined
from the indirect measurement, up to uncertainties in the density
profile and, for the signal from the outer regions, uncertainties in
the properties of substructure.  These uncertainties in the dark
matter distribution enter into $\Phi_{\rm A}(\psi_{\rm cross})$ and
$\Phi_{\rm D}(\psi_{\rm cross})$.

In the following sections we discuss in detail the approach outlined
here for the case of gamma-ray observations of dwarf galaxies.  We
illustrate the proposed method for selected dwarfs, show the expected
signals for several example benchmark dark matter scenarios, and
indicate the range of dark matter particle parameters for which a
transition between annihilation and decay would occur in these objects.

\section{Dwarf galaxies}
\label{dwarfs}

Dwarf galaxies are extremely dark-matter--dominated, with
mass-to-light ratios in the range $100 \, M_\odot / L_\odot < M/L <
1000 \, M_\odot / L_\odot$~\cite{Simon:2007dq, Martin:2008wj}.  High
dark matter densities coupled with minimal foregrounds due to a
scarcity of astrophysical gamma-ray sources make these objects
excellent targets for indirect dark matter searches in GeV and TeV
gamma-rays~\cite{Baltz:2008wd, dG, Evans:2003sc, Strigari:2007at,
  Essig:2009jx, MAGICdG, HESSdG, Driscoll:2008zz, Wagner:2009wp,
  Abdo:2010ex}.  The predicted emission from dark matter decay or
annihilation in Milky Way dwarfs has a large angular extent ($\sim$
few degrees), which makes it possible to map the angular distribution
of an observed signal.

We illustrate the proposed technique for three Milky Way dwarf
galaxies: Draco, Ursa Minor, and Sagittarius.  These dwarf galaxies
are among the most optimistic for detection in gamma-rays
(e.g., Refs.~\cite{Essig:2009jx, Abdo:2010ex}), and are all accessible
targets for dark matter searches with HESS, VERITAS, or MAGIC\@.  It
is important to note that Sagittarius is undergoing tidal disruption
by the Milky Way which induces substantial uncertainties in the
structural properties of this dwarf galaxy, and in turn in the
predicted annihilation and decay signals.  However, due to its close
proximity and inferred large mass-to-light ratio Sagittarius is often
considered an excellent target for dark matter searches in gamma-rays,
so for this reason we include it here.

The dominant gamma-ray signal from dark matter annihilation or decay
in dwarf galaxies of the Milky Way is produced in conjunction with the
hadronization, fragmentation, and subsequent decay of the Standard
Model particles in the final states.  In channels with charged
leptons in the final states, internal bremsstrahlung gamma-rays are
also generated.  In addition, energetic electrons and positrons
produced in these processes give rise to secondary photons at various
wavelengths via inverse Compton scattering off the ambient photon
background and synchrotron emission due to the presence of magnetic
fields.  Since both the magnetic field and starlight density in dwarf
galaxies are small, the dominant energy loss mechanism for electrons
and positrons at GeV--TeV energies is expected to come from the
upscattering of Cosmic Microwave Background photons, resulting in
secondary photons with significantly smaller energies than the primary 
gamma-ray emission from these channels~\cite{Baltz:2004bb,
  Jeltema:2008ax, Kuhlen:2009jv, Kistler:2009xf}.  Thus, this secondary
emission is only significant at the lowest part of the energy ranges
considered here and only for some channels, so for simplicity we
ignore this contribution.

Substructure in the halo of a Milky Way-sized galaxy, including
multiple generations of nested sub-subhalos, is resolved in numerical
simulations, including subhalos with properties matching those of
known dwarf galaxies~\cite{Diemand:2008in, Springel:2008cc,MWsats}.
Simulations indicate that the dark matter density profiles of
dwarf--galaxy--sized satellites can be described by the canonical
density profiles used to describe larger halos, although some
uncertainties remain since tidal disruption may lead to mass
redistribution, e.g., significant mass loss in the outer regions of
the halo~\cite{Hayashi:2002qv, Kazantzidis:2003hb, Penarrubia:2007zx,
  Walker:2009zp}.  The properties of substructure in dwarf galaxy
halos are even less certain, but for completeness we consider this
possibility as well.

The dwarf's halo and its subhalos were likely formed before tidal
stripping of the dwarf halo took place.  Since studies suggest that
most of the tidal mass-loss occurs in the outer regions of the halo,
leaving the inner regions largely intact, we set the properties of the
halos and subhalos before any tidal effects, and only account for
tidal effects by truncating the dwarf halo at a radius $r_{\rm cut}$
(c.f.~\S\ref{smooth}) when performing the line-of-sight integral for
the intensity calculations.  In essence, we assume that all of the
mass, smooth and subhalos, outside of $r_{\rm cut}$ has been removed
by tidal stripping, while everything within $r_{\rm cut}$ remains
unchanged.

We treat separately the contributions from the smooth halo and
substructure components to the gamma-ray signal.  The smooth halo case
alone provides a lower limit on the gamma-ray signal from annihilation
for our assumed density profile and represents the steepest angular
emission profile.  On the other hand, simulations indicate that a
scaled-down host subhalo population represents the maximum expected
abundance of sub-substructure~\cite{Diemand:2008in, Springel:2008cc},
so we model the subhalo population of each dwarf in this way to
consider the upper limits on the total annihilation flux and on the
shallowness of the angular emission profile in the annihilation case.

\subsection{The smooth halo}
\label{smooth}

We describe the mass distribution of the smooth dark matter halo of
each dwarf galaxy by a Navarro, Frenk and White (NFW) density
profile~\cite{NFW} 
\begin{equation}
\label{eq:NFW}
 \rho_{\rm sm}(r)=\frac{\rho_{\rm s}}{r/r_{\rm s} \left(1+r/r_{\rm
     s}\right)^2}~, 
\end{equation}
where $r$ is the distance to the center of the object, $\rho_{\rm s}$
is a characteristic density, and $r_{\rm s}$ is a scale radius.  To
account for tidal stripping of the outer regions of the dwarf's halo
we truncate the profile at a radius $r_{\rm cut}$ given by the
Roche-limit criterion
\begin{equation}
\label{eq:roche}
r_{\rm cut} \simeq \left( \frac{G M_{\rm dwarf} d_{\rm GC}^{2}}{2
  \sigma^{2}_{\rm MW}} \right)^{1/3}~, 
\end{equation}
where $M_{\rm dwarf}$ is the mass of the dwarf's halo used here to
determine the tidal radius, $d_{\rm GC}$ is its distance to the
Galactic Center, and $\sigma_{\rm MW}$ is the velocity dispersion of
the Milky Way at the satellite's position.  Following
Ref.~\cite{Strigari:2008ib}, we conservatively adopt $M_{\rm
  dwarf}=10^{9}$ M$_{\odot}$, rather than the virial mass $M_{\rm
  halo}$ implied by each dwarf's structural parameters
(c.f. \S\ref{substructure}), and $\sigma_{\rm MW}=200$~km/s to
determine the tidal radius of each dwarf's halo.  However, the virial
halo mass $M_{\rm halo}$, not $M_{\rm dwarf}$, is used in the
following sections to set the structural properties of each dwarf halo
and its subhalo population.  Measured and derived properties of the
dwarf galaxies considered here are summarized in
Table~\ref{tab:dwarfparams}.

\begin{table}
\caption{\label{tab:dwarfparams}Measured and derived properties of
  selected dwarf galaxies.  Heliocentric distance $d_{\odot}$ is taken
  to be the distance to the center of the object.  The values of
  $r_{\rm s}$ and $v_{\rm max}$ are taken from Ref.~\cite{Walker:2009zp},
  and following Ref.~\cite{Evans:2003sc}, we adopt the same values for
  these parameters for Sagittarius as for Draco.  The scale density
  $\rho_{\rm s}$ is derived from $r_{\rm s}$ and $v_{\rm max}$
  assuming a NFW density profile, and the tidal radius $r_{\rm cut}$
  is calculated from Eq.~\ref{eq:roche}.  See text for details. \\} 
\begin{tabular}{cccccccc}
\hline \hline
Object & $d_{\odot}$ & $d_{\rm GC}$ & $r_{\rm s}$ & $v_{\rm max}$ &
$\rho_{\rm s}$ & $r_{\rm cut}$ & Refs.\\ 
 & (kpc) & (kpc) & (kpc) & (km/s) & (M$_{\odot}$ kpc$^{-3}$) & (kpc) & \\
\hline 
Draco & 76 & 76 & 0.795 & 22 & $6.6 \times 10^{7}$ & 6.8  &
\cite{Bonanos:2003hd, Mateo:1998wg}\\ 
Ursa Minor (UMi) & 66 & 68 & 0.795 & 21 & $6.0 \times 10^{7}$  & 6.3 &
\cite{Mateo:1998wg}\\ 
Sagittarius (Sgr) & 24 & 16 & 0.795 & 22 & $6.6 \times 10^{7}$ & 2.4 &
\cite{Mateo:1998wg}\\ 
\hline \hline
\end{tabular}
\end{table}

Now, writing explicitly the factor $\Phi$ in
Eq.~\ref{eq:intenscompact}, the intensity $I_{\rm sm}$ from dark 
matter annihilation or decay in the smooth halo from an angle $\psi$
between the line-of-sight direction and the center of the object is
\begin{equation}
\label{eq:ismooth}
I_{{\rm sm},x}(\psi)=\frac{P_{x}}{4\pi} \int_{los} ds \, \rho_{\rm
  sm}^{i}(r(s,\psi))~. 
\end{equation}
For decay $(i,x)=(1,{\rm D})$ and for annihilation $(i,x)=(2,{\rm
  A})$.  For observations of the inner regions of Milky Way dwarf
galaxies we take $\psi=R/d_{\rm GC}$ since $R \ll d_{\rm GC}$, where
$R$ is the projected radius from the center of the dwarf.  The
line-of-sight integral extends from $s=-z_{\rm max}$ to $s=+z_{\rm
  max}$ where $z_{\rm max}=\sqrt{r_{\rm max}^{2} - R^{2}}$, and here
we take $r_{\rm max}=r_{\rm cut}$, given by Eq.~\ref{eq:roche}.  The
dark matter particle properties are encoded in the factors $P_{\rm D}$ 
and $P_{\rm A}$ as defined in Eq.~\ref{eq:pfactors}.

\subsection{Substructure}
\label{substructure}

We consider collectively the emission from subhalos within the
dwarf galaxy halo, summing over the contribution to the gamma-ray
signal from subhalos of all masses.  We assume that the density
profile of each subhalo can be described by a NFW profile.  The
individual subhalo differential luminosity $L$ (photons per time per
energy) as a function of subhalo mass $M_{\rm sub}$ is proportional to
the integral over the subhalo volume of $\rho$ in the case of decay or
$\rho^{2}$ in the case of annihilation: 
\begin{eqnarray}
L_{\rm D} & = & P_{\rm D}  \int \,{\rm d}V_{\rm sub} \rho_{\rm sub}
\propto  M_{\rm sub} \\ 
L_{\rm A} & = & P_{\rm A} \int \,{\rm d}V_{\rm sub} \rho^{2}_{\rm sub}
\propto  M_{\rm sub}\,\frac{c^3}{f^2(c)}~. 
\label{eq:lsub}
\end{eqnarray}
The subhalo concentration $c=r_{\rm vir}/r_{\rm s}$ where $r_{\rm
  vir}$ is the virial radius of the subhalo, corresponding to an
average enclosed overdensity of 200 times the critical density, and
$f(c)=\ln (1+c) + (c/(1+c))$.

We adopt a power-law mass function for the subhalos, ${\rm d}N/{\rm d}
M \propto M^{-\alpha}$ with $\alpha = 1.9$ and minimum and maximum
subhalo mass $M_{\rm min}=10^{-6}$ M$_{\odot}$ and $M_{\rm max}=M_{\rm
halo}$, where $M_{\rm halo}$ is the virial mass of the dwarf halo
determined by the density profile parameters given in
Table~\ref{tab:dwarfparams}, without tidal stripping, within a virial 
radius $r_{\rm vir}$.  

\begin{figure*}[t] 
   \centering
   \includegraphics[width=1.\textwidth]{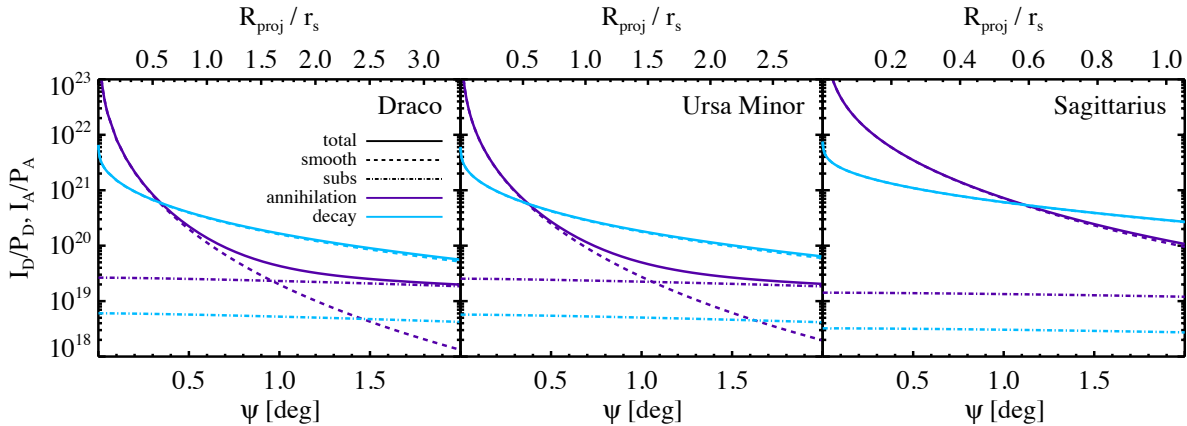} 
   \caption{Dependence of the intensity from decay ({\it blue})
     and annihilation ({\it purple}) on line-of-sight direction
     $\psi$ from the center of the object for selected dwarf galaxies.
     The contributions from the smooth halo ({\it dashed}),
     substructure ({\it dot-dashed}), and the total ({\it solid}) are
     shown.  The corresponding projected radius $R_{\rm proj}$ in
     units of the halo scale radius $r_{\rm s}$ is labeled on the top
     axis.  For generality, the amplitudes of the curves for decay and
     annihilation have been scaled by the factors $P_{\rm D}$ and 
     $P_{\rm A}$ respectively, which depend on the assumed particle
     properties.  $I_{\rm D}/P_{\rm D}$ is shown in units of GeV
     cm$^{-2}$ sr$^{-1}$, and $I_{\rm A}/P_{\rm A}$ is shown in units
     of GeV$^{2}$ cm$^{-5}$ sr$^{-1}$.
\label{fig:iprofiles}}
\end{figure*}

We note that although subhalo masses are described by the quantity 
$M_{\rm sub}$ for the purpose of defining a mass-concentration relation,
in numerical simulations subhalos are typically identified by measuring
the parameters $v_{\rm max}$, the maximum circular velocity of particles
in the subhalo, and $r_{v_{\rm max}}$, the radius at which that
maximum occurs.  These two parameters are sufficient to specify a NFW
density profile for the subhalo and thereby fix the structural
properties.  The quantity $M_{\rm sub}$ is the mass within the virial
radius of a subhalo with a given NFW density profile, with the virial
radius as defined previously.  It is important to keep in mind that
subhalos may suffer mass loss as a result of tidal stripping, and thus
$M_{\rm sub}$ may not accurately reflect the current bound mass of the
subhalo.  As in prior work (e.g., Ref.~\cite{Kuhlen:2008aw}) based on
the results of simulations, $M_{\rm sub}$ is used here in conjunction
with the concentration parameter $c$ to parameterize the density
profile of the subhalos and determine a $L_{\rm A}$-$M_{\rm sub}$
relation.  Although the possible impact of tidal stripping, in
particular the corresponding decrease of the bound subhalo mass, on
the annihilation luminosity is expected to be small since most
annihilation occurs in the inner regions of the subhalo, the effect on
the decay luminosity could be greater.  However, as the decay signal
from substructure is subdominant with respect to the smooth halo decay
signal at all radii we consider (see Fig.~\ref{fig:iprofiles}), we do
not attempt to account for this uncertain effect here.

The mass-luminosity relation for annihilation can be
approximated by $L_{\rm A}  \propto M_{\rm sub}^{\beta}$ with
$\beta=0.87$ (0.94) for $M_{\rm sub}$ greater than (less than)
$10^{6}$ M$_{\odot}$.  The value of $\beta$ is determined from
Eq.~\ref{eq:lsub} using the subhalo mass-concentration relation
from Ref.~\cite{Kuhlen:2008aw}, i.e., $c_{\rm sub}(M_{\rm sub}) \propto
M_{\rm sub}^{\delta}$ with $\delta = -0.06$ (-0.025) for $M$ greater than
(less than) $10^{6}$ M$_{\odot}$.

The particle physics factor can be separated from the contribution to
the luminosity depending on the subhalo density profile by defining
$\mathcal{L}_{x}=L_{x}/P_{x}$, and hence we can write
\begin{equation}
\frac{{\rm d}N}{{\rm d}\mathcal{L}}=\frac{{\rm d}N}{{\rm
    d}M}\frac{{\rm d}M}{{\rm d}\mathcal{L}}~. 
\end{equation}
The mass function is normalized such that 
$\int_{M_{\rm min}}^{M_{\rm max}} {\rm d}N/{\rm d}M = 1$, 
so integrating the individual subhalo $\mathcal{L}_{x}$ over
the mass function gives the average contribution to the decay or
annihilation rate from a single subhalo 
\begin{eqnarray}
\label{eq:lsubs}
\mathcal{L}_{_{\rm subs, D}}&=&\frac{f_{\rm subs} M_{\rm halo}}{N_{\rm subs}}\\
\mathcal{L}_{_{\rm subs, A}}&=&\int_{\mathcal{L}_{_{\rm min,
      A}}}^{^{\mathcal{L}_{\rm max, A}}} \!\!\! {\rm
  d}\mathcal{L}_{\rm A}\, \mathcal{L}_{\rm A}\, \frac{{\rm d}N}{{\rm
    d}\mathcal{L}_{\rm A}}
\end{eqnarray}
where $f_{\rm subs}$ is the mass fraction in substructure and
$\mathcal{L}_{\rm min, A}$ and $\mathcal{L}_{\rm max, A}$ are the
values of $\mathcal{L}_{\rm A}$ corresponding to $M_{\rm min}$ and
$M_{\rm max}$, respectively.

The intensity in a direction $\psi$ from subhalos, analogous to the
smooth halo intensity given in Eq.~\ref{eq:ismooth}, is then given by 
\begin{equation}
I_{{\rm subs},x}(\psi)=\frac{P_{x}}{4\pi} \int_{los} ds \,
\mathcal{L}_{_{{\rm subs},x}}\, n_{_{\rm subs}}(r(s,\psi))~, 
\end{equation}
where $n_{_{\rm subs}}(r)$ is the number density of subhalos at a
radius $r$ from the halo center. 

Following the results of recent numerical simulations we describe the
radial distribution of subhalos with an Einasto
profile~\cite{Einasto}: 
\begin{equation}
\label{eq:einsub}
n_{_{\rm subs}}(r) \propto \exp\left\{-\frac{2}{\alpha_{\rm Ein}}
\left[ \left( \frac{r}{r_{-2}} \right)^{\alpha_{\rm Ein} } -1 \right]
\right\}~, 
\end{equation}
with $\alpha_{\rm Ein}=0.68$ and $r_{-2}=0.81 \, r_{\rm vir}$, as in
Ref.~\cite{Springel:2008cc}.  The subhalo number density is normalized
so that 10\% of the mass of the smooth halo is condensed into subhalos
with masses between $10^{-5}$ and $10^{-2}$ $M_{\rm halo}$.  For the
dwarf galaxies considered here, this results in $\sim 30\%$ of $M_{\rm
  halo}$ bound in substructure.

The angular dependence of the gamma-ray intensity from annihilation
and decay is shown in Fig.~\ref{fig:iprofiles} for our three example
dwarf galaxies.  The contributions from substructure and the smooth
halo are shown separately, along with the total of these signals from
each process.  Note that we have not subtracted the subhalo mass
density from the smooth halo in determining the signals from
annihilation and decay in the smooth halo, so for scenarios with
substructure, this leads to a slight overestimate of the smooth halo
signal and correspondingly, the total signal. This is a negligible
effect for annihilation since the correction to the mass density applies
preferentially to the outer regions of the halo where the subhalos
represent a larger mass fraction, and where the smooth halo
annihilation signal is small and subdominant to the substructure
annihilation signal.  Likewise, this correction is a small effect for
decay, so for simplicity we do not account for it here.  The
contribution from annihilation or decay in substructure (blue and
purple dot-dashed curves) tends to be nearly parallel to the smooth
halo contribution in the case of decay (blue dashed curves) at angles
$\gtrsim 1^{\circ}$.  This is expected, as the number density of
subhalos scales approximately as $\rho_{\rm sm}$ outside of the inner
region of the halo, as the dark matter decay rate in the smooth
halo does.  Note that decay in substructure is always subdominant
relative to decay in the smooth halo, even in the maximal substructure
scenario we consider here.  The shapes of the curves are quite similar
for all three objects, however the amplitude of the curves for
Sagittarius is somewhat larger, reflecting the fact that this dwarf
galaxy has similar structural properties to the other two but is at a
significantly smaller distance.

\section{Gamma-ray spectra from dark matter annihilation and decay}
\label{spectra}

With the goal of presenting the results in a general and
model-independent way, we consider here the generic features in the
gamma-ray spectrum implied by dark matter annihilations and decays,
whose products we will assume to be two Standard Model
particles.  Cases with additional SM particles in the final state and
more exotic scenarios have also been considered~\cite{extramodels,
  Meade:2009iu, Papucci:2009gd}.  An investigation of the use of our
proposed method in those cases would require additional dedicated
analyses.

The spin of the state constituted by the s-wave of two
non-relativistic dark matter particles can only have integer values 
(if it is a WIMP, it can only be 0, 1 or 2).  Thus, annihilation of
dark matter particles with mass $m_{\chi}$ can be described as the
decay of the s-wave of a state with mass 2$m_{\chi}$ and integer
spin~\cite{Cirelli:2008pk}.  Hence, the possible final states (at tree
level) with only two Standard Model particles produced from dark matter
annihilations are: $W^+W^-$, $ZZ$, $Zh$, $hh$, $l^+l^-$ and
$q\bar{q}$, where $l$ and $q$ represent any lepton or quark,
respectively.

On the other hand, the allowed channels from dark matter decay are
much less constrained and, in principle, the spin of the initial state
can be integer or half-integer. Although mixed channels involving a
Standard Model particle and a new particle or three-body decays are
possible, we will not consider these possibilities and instead
restrict our study to the case of only two Standard Model particles in
the final state. In the case of decays, in addition to the allowed
channels for annihilation, which correspond to scalar dark matter
decay, there is the possibility of semileptonic channels, such as
$W^+l^-$, if dark matter is a fermion.

\begin{figure*}[t]
   \centering
   \includegraphics[width=1.\textwidth,clip=true]{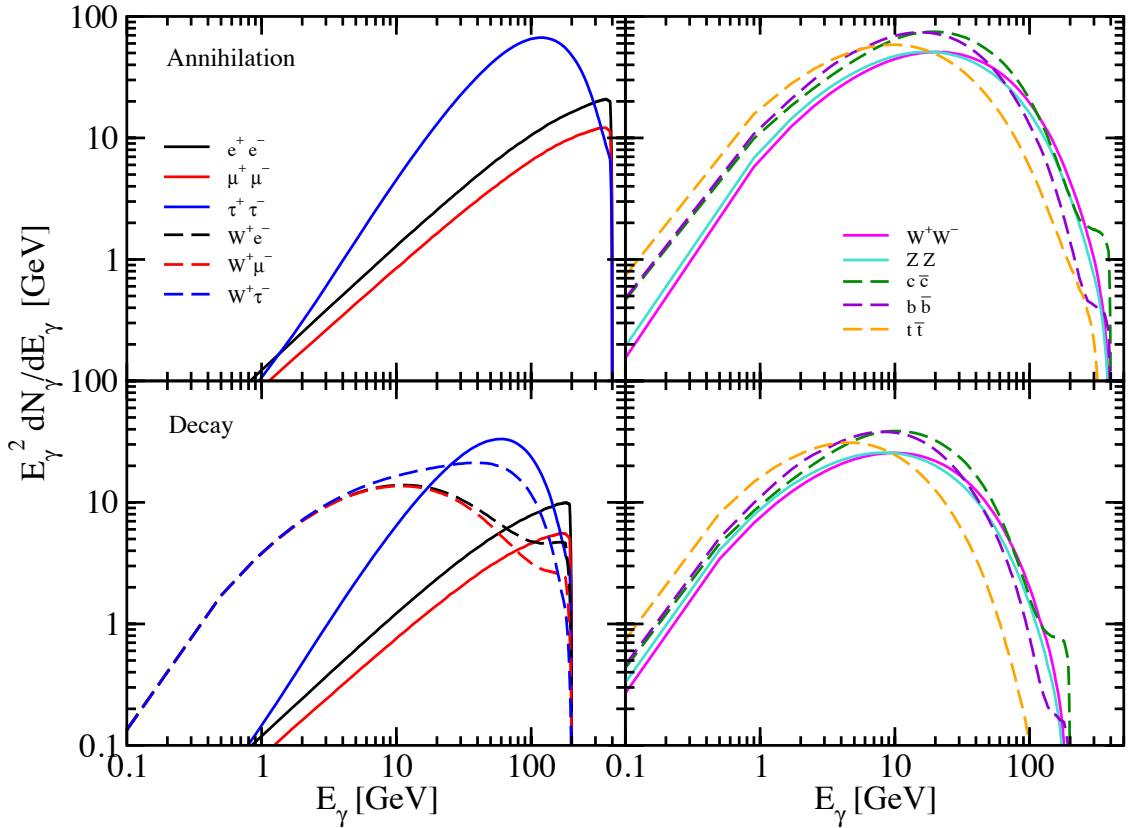}
   \caption{Energy spectra ($E_\gamma^2 \, \, {\rm d}N_\gamma/{\rm
       d}E_\gamma$) for different possible final states with two
     Standard Model particles for a 400~GeV dark matter candidate.  In
     the upper panels we show the case of annihilation and in the
     lower panels that of decay.  {\it Left panels:} leptonic and
     semileptonic channels, as labeled. Note that the semileptonic
     channels ($W^+e^-$, $W^+\mu^-$, and $W^+\tau^-$) can only be
     present for decay.  {\it Right panels:} hadronic and gauge boson
     channels, as labeled.
\label{fig:spectra}}
\end{figure*}

We simulate the hadronization, fragmentation and decay of different
final states with the event generator PYTHIA
6.4~\cite{Sjostrand:2006za}.  For each mass and channel we consider
$10^7$ events distributed in 500 logarithmically-spaced energy bins.
In Fig.~\ref{fig:spectra} we show the gamma-ray spectra for a 400~GeV
dark matter particle for annihilation (upper panels) and decay (lower
panels).  In the left panels we show the leptonic and semileptonic
(decay-only) channels, whereas in the right panel we show the hadronic
and gauge boson channels.  It is clear from the figure that leptonic
channels typically give rise to a harder spectrum.  Indeed, in the
case of leptonic channels (in particular $e^+e^-$ and $\mu^+\mu^-$), the
cutoff would be very sharp around a maximum energy. We emphasize,
however, that the presence of this cutoff in the spectrum would not
allow one to determine the dark matter mass unless one knows if the
signal comes from annihilation or decay, due to the factor of 2
difference in the cutoff energy between these processes.  
 
As it can be seen in the figure, one could generically classify the
spectra associated with these final states as either soft (hadronic or
gauge boson) or hard (leptonic), with semileptonic channels being a
mixture of the two.  Thus, for the sake of simplicity when showing
explicit examples, we will consider combinations of soft-hard or
hard-soft channels for annihilation-decay, which in general represent
the extreme cases. We comment that although it is possible for the
dominant final states for both annihilation and decay to be the same
(which would yield similarly shaped energy spectra), the maximum
energy of the photons from dark matter annihilation is twice the
energy of those from dark matter decay. Thus, as can be seen from 
Fig.~\ref{fig:spectra}, in general the
gamma-ray spectra from dark matter annihilation and decay are
different~\footnote{In the case of dark matter annihilation dominantly
  into top quarks and decay dominantly into any of the other hadronic or
  gauge boson channels shown in Fig.~\ref{fig:spectra}, the
  distinction, on the basis of the energy spectra alone, might be  
  challenging.}.

\section{Results}
\label{results}

The aim of this work is to propose an observing strategy to maximize
the information that could be obtained from gamma-ray experiments in
the case that a signal from dark matter is clearly detected.  Our goal
is not to evaluate the detectability of specific scenarios with
current instruments; for the interested reader we reference some of
the many studies that have carefully addressed that subject with
observations of dwarf galaxies~\cite{MAGICdG, HESSdG, Driscoll:2008zz,
  Wagner:2009wp, Abdo:2010ex}.  However, let us emphasize that in
order for this approach to be applicable with current experiments, a
signal just beyond the limits established so far must be discovered to
yield sufficient statistics to map the signal and determine the energy
spectrum over an extended angular region.

The first requirement in order to use this technique is that the source 
is resolved as an extended source.  In particular, we assume that the
signal can be binned into several annuli centered on the source.  This
is in principle possible with the angular resolution of current
experiments ($\sim 0.1^{\circ}$ at the relevant energies) for
observations of dwarf spheroidal galaxies, since the angular extent of 
the predicted dark matter signal is as large as $\sim$ few degrees.
In addition, this technique requires that the signal in each annulus
is detected with sufficient statistics to reconstruct the energy
spectrum.  In the following we proceed under the assumption that a
signal meeting these conditions has been detected.

\begin{figure*}[t] 
   \centering
   \includegraphics[width=1.\textwidth]{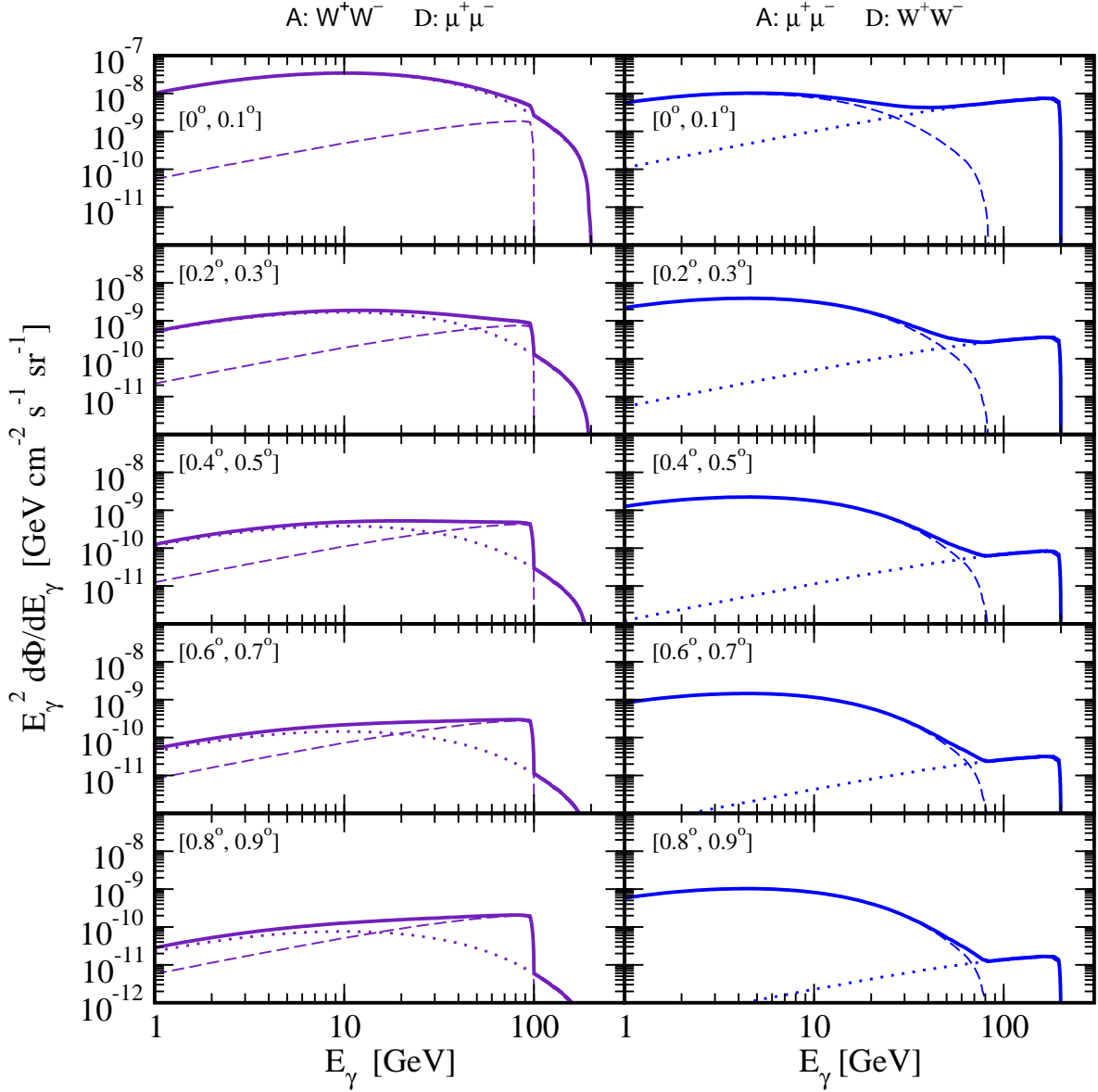}
   \caption{Energy spectra in different annuli centered on Draco for
     $m_\chi=$~200~GeV and for two combinations of channels.  {\sl Left
     panels:} Annihilation into $W^+ W^-$ (soft) and decay into $\mu^+
     \mu^-$ (hard).  {\sl Right panels:} Annihilation into $\mu^+
     \mu^-$ (hard) and decay into $W^+ W^-$ (soft).  Dotted lines
     represent the contribution from annihilation, dashed lines that
     from decay, and solid lines the sum of the two. In each column,
     the spectra for alternating annuli of $0.1^{\circ}$ width is
     shown, with the innermost annuli in the top panels. The
     contribution from substructure is included and here we take
     $\langle \sigma v \rangle = 3 \times 10^{-26}$~cm$^3$~s$^{-1}$
     and $\tau_\chi = 10^{29}$~s.  The change in the shape and
     amplitude of the total ({\it thick solid lines}) from inner to
     outer annuli indicates that both annihilation and decay are
     present.
\label{fig:annuli}} 
\end{figure*}

In Fig.~\ref{fig:annuli} we illustrate the proposed method for a
scenario in which both annihilation and decay contribute appreciably
to the observed signal from the Draco dwarf galaxy by showing the
energy spectrum as a function of the angle from the center of the
object.  The energy spectrum in alternating annuli of $0.1^{\circ}$
width centered on Draco is shown out to an angular radius of
$0.9^{\circ}$ (from top to bottom) for a dark matter particle mass of 
$m_\chi=$~200~GeV.  Two combinations of channels are shown.  The left
(right) column shows the case of annihilation into a soft (hard)
channel and decay into a hard (soft) one.  The channels $\mu^+ \mu^-$
and $W^+ W^-$ have been chosen as representative of hard and soft
channels, respectively.  In each panel, dashed lines represent the
contribution from decay, dotted lines represent that from annihilation,
and the thick solid lines represent the total contribution.  We have
taken $\langle \sigma v \rangle = 3 \times 10^{-26}$~cm$^3$~s$^{-1}$
and $\tau_\chi = 10^{29}$~s.  Note that although we have included the
contribution of substructure, it is a subdominant effect for both
annihilation and decay for the annuli considered in this figure (see
Fig.~\ref{fig:iprofiles}).  The presence of substructure increases the
contribution from annihilation primarily in larger annuli than shown
in this figure.  As expected, a significant change in the spectrum is
clearly seen in Fig.~\ref{fig:annuli} for both combinations of
channels at around $E = m_{\chi}/2$, i.e., the maximum energy for
photons from dark matter decay.  The spectral change is a signature of
both annihilation and decay contributing significantly to the signal.

\begin{figure*}[t] 
   \centering
   \includegraphics[width=1.\textwidth]{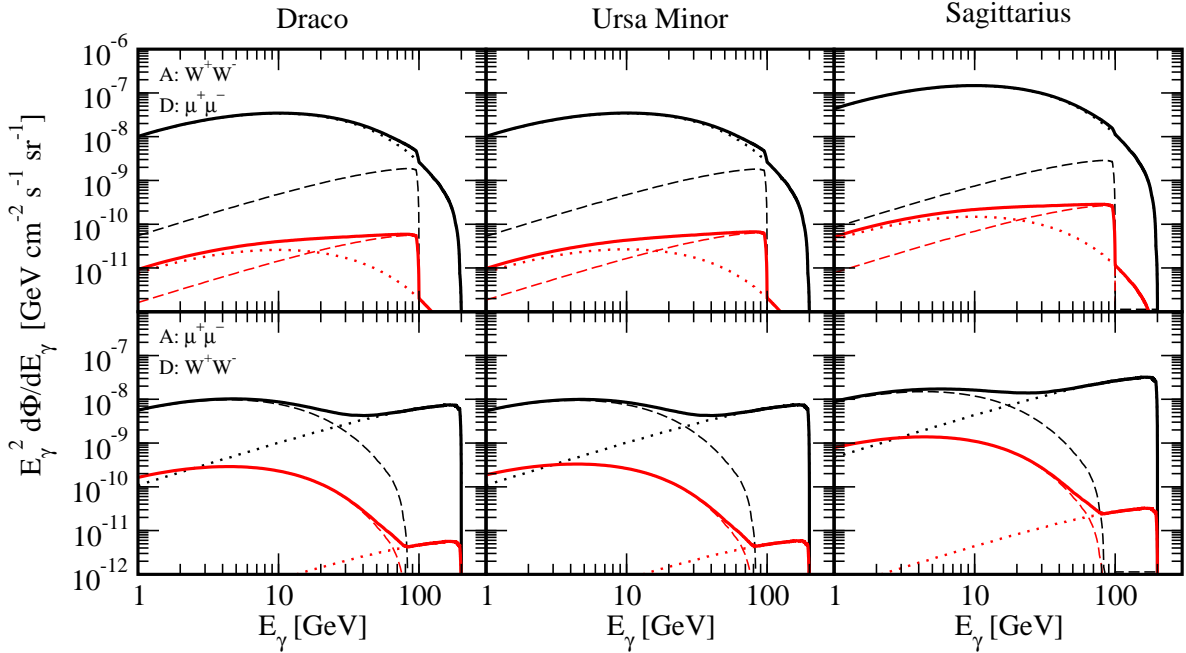} 
   \caption{Energy spectra for $m_\chi=$~200~GeV from the contribution
     of the innermost annuli ({\it black lines},
     $[0.0^{\circ},0.1^{\circ}]$) and the outermost annuli ({\it red
       lines}, $[1.9^{\circ},2.0^{\circ}]$) we consider, for three
     dwarf galaxies.  We show in the upper and lower panels the same
     two combinations of channels  as in Fig.~\ref{fig:annuli},
     adopting the same values for $\langle \sigma v \rangle$ and
     $\tau_\chi$ and again including substructure.  As in
     Fig.~\ref{fig:annuli}, dotted lines indicate the contribution
     from annihilation, dashed lines that from decay, and the thick
     solid lines show the total signal. 
\label{fig:inout}}
\end{figure*}

As a summary of the discussion above, in Fig.~\ref{fig:inout} we show
the contributions, again including substructure, from the innermost
(black lines, $[0.0^{\circ},0.1^{\circ}]$) and the outermost (red lines,
$[1.9^{\circ},2.0^{\circ}]$) annuli we consider, for
$m_\chi=$~200~GeV. We show the results for the three studied dwarf
galaxies for the same two combinations of channels as in
Fig.~\ref{fig:annuli}, as indicated, and the same values for the
annihilation cross section and lifetime are used as in
Fig.~\ref{fig:annuli}.  The changes in spectral shape and amplitude
between the innermost and outermost annuli are easily identified.

\begin{figure*}[t] 
   \centering
   \includegraphics[width=1.\textwidth]{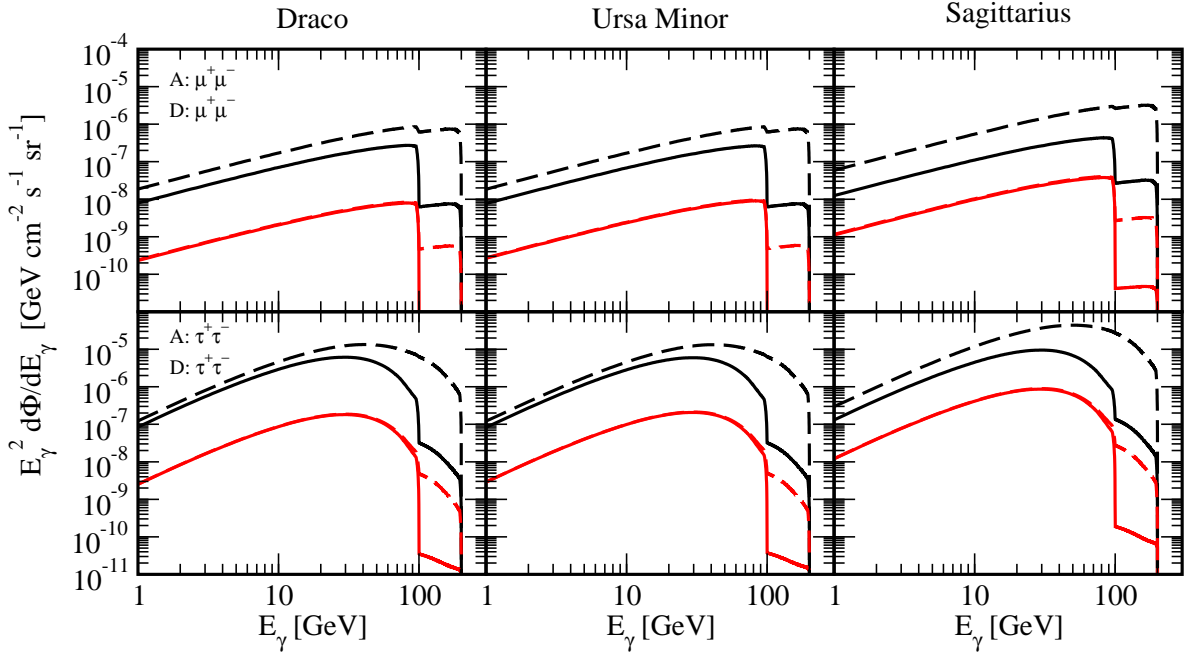} 
   \caption{Energy spectra for $m_\chi=$~200~GeV from the contribution
     of the innermost annuli ({\it black lines},
     $[0.0^{\circ},0.1^{\circ}]$) and the outermost annuli we consider 
     ({\it red lines}, $[1.9^{\circ},2.0^{\circ}]$), for the three
     studied dwarf galaxies. In each row we assume dark matter
     annihilates and decays dominantly into the same channel and
     depict the total contribution. {\it Upper panels:} $\mu^+\mu^-$,
     $\tau_\chi = 7 \times 10^{26}$~s and $\langle \sigma v \rangle =
     3 \times 10^{-26}$~cm$^3$~s$^{-1}$ ($\langle \sigma v \rangle = 3
     \times 10^{-24}$~cm$^3$~s$^{-1}$) for the solid ({\it dashed})
     lines. {\it Lower panels:} $\tau^+\tau^-$, $\tau_\chi = 2 \times
     10^{26}$~s and $\langle \sigma v \rangle = 3 \times
     10^{-26}$~cm$^3$~s$^{-1}$ ($\langle \sigma v \rangle = 6 \times
     10^{-24}$~cm$^3$~s$^{-1}$) for the solid ({\it dashed})
     lines. These four cases fit the PAMELA data in decay-only or
     annihilation-only scenarios~\cite{Meade:2009iu, Cirelli:2009dv}.
     Note that the parameters are chosen so that the total signal
     would also fit the data.
\label{fig:inoutPAMELA}}
\end{figure*}

On the other hand, we also show in Fig.~\ref{fig:inoutPAMELA} an
example for two dark matter models that could potentially explain the
PAMELA data: one in terms of decay as the primarily contributor plus a
negligible contribution to the PAMELA signal from annihilation with a
thermal cross section (solid lines), and one in which decay and
annihilation each contribute similarly to the PAMELA signal (dashed
lines).  We depict the the total contribution for $m_\chi=$~200~GeV
from dark matter annihilation plus decay for the three dwarf galaxies
and for two channels: $\mu^+\mu^-$ (top row) and $\tau^+\tau^-$
(bottom row).  In each row, the same channel is assumed for both
annihilation and decay rather than combinations of channels as in
Figs.~\ref{fig:annuli} and~\ref{fig:inout}.  For the $\mu^+\mu^-$
case, $\tau_\chi = 7 \times 10^{26}$~s and $\langle \sigma v \rangle =
3 \times 10^{-26}$~cm$^3$~s$^{-1}$ for the dominantly-decay scenario
(solid lines) while $\langle \sigma v \rangle = 3 \times
10^{-24}$~cm$^3$~s$^{-1}$ for the two-process scenario (dashed lines).
For the $\tau^+\tau^-$ case, $\tau_\chi = 2 \times 10^{26}$~s and
$\langle \sigma v \rangle = 3 \times 10^{-26}$~cm$^3$~s$^{-1}$
($\langle \sigma v \rangle = 6 \times 10^{-24}$~cm$^3$~s$^{-1}$) for
the solid lines (dashed lines).  In each panel, black and red lines
represent the result for the innermost ($[0^{\circ},0.1^{\circ}]$) and
outermost ($[1.9^{\circ},2^{\circ}]$) annulus, respectively.  These
four cases are found to fit the PAMELA data in the case of decay-only
and annihilation-only~\cite{Meade:2009iu, Cirelli:2009dv}, but we have
chosen the parameters so that the combination of the two types of signal
would also fit the data.  We see that very different spectra are
obtained from the center of the galaxies as compared to regions
further away, which is a clear signature of the presence of dark
matter decay as well as annihilation.

\begin{figure*}[t] 
   \centering
   \includegraphics[width=0.85\textwidth]{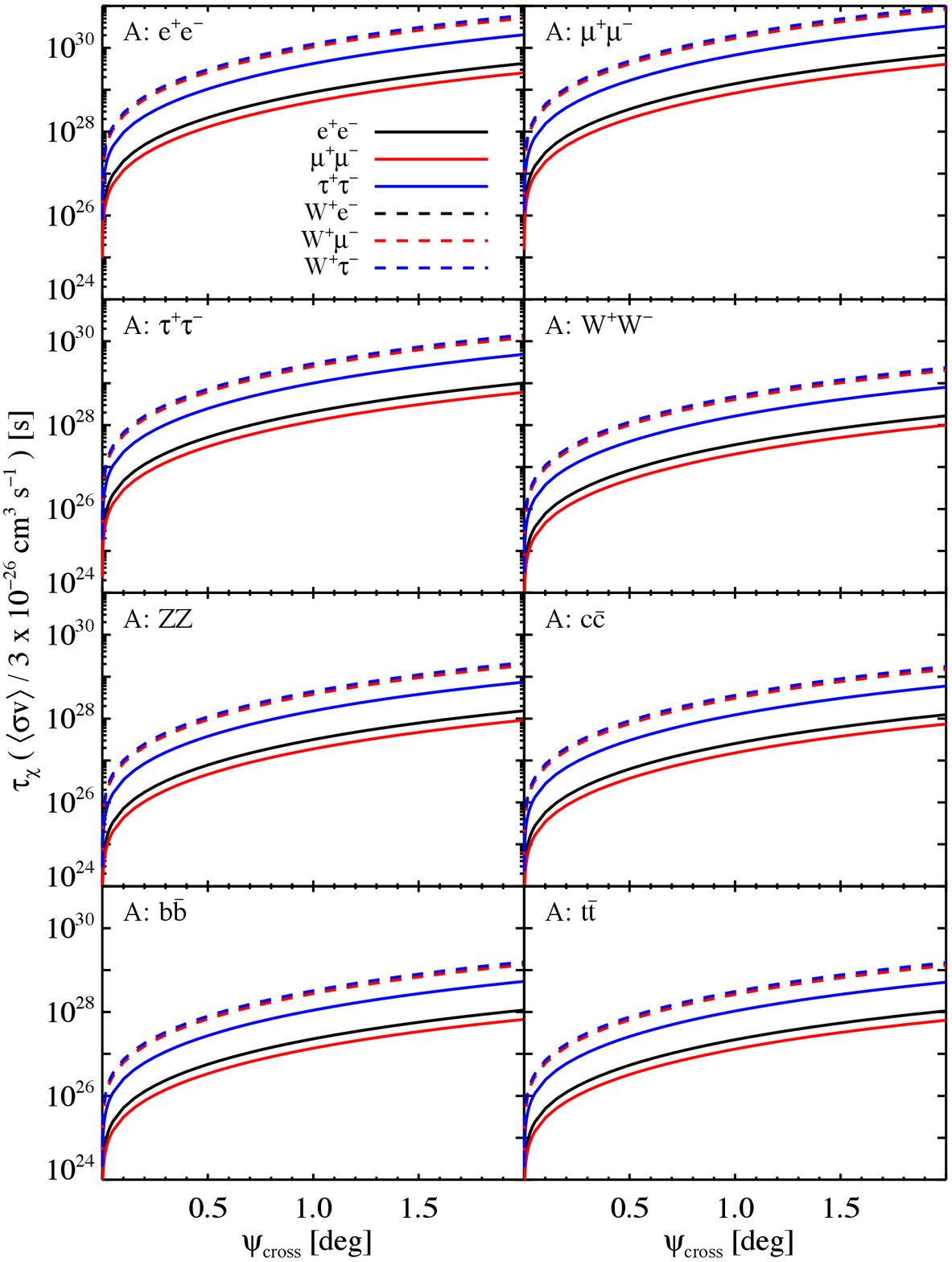}  
   \caption{Lifetime $\tau_\chi$ at which the intensities from
   annihilation and decay for $E > 1$~GeV are equal at an observation 
   angle $\psi_{\rm cross}$ from the center of the dwarf galaxy, for
   Draco, without substructure.  The dark matter particle mass is
   $m_{\chi}=200$~GeV.  Each panel shows curves for a single
   annihilation channel, assuming decay into leptonic or semileptonic
   channels (as labeled). The red dashed curves for the $W^{+}\mu^{-}$
   decay channel fall on top of the black dashed curves for the
   $W^{+}e^{-}$ decay channel because the photon yields above 1~GeV
   for these two decay channels for this particle mass are almost
   identical.  Smaller values of the lifetime for a fixed annihilation
   cross-section correspond to smaller values of $\psi_{\rm
     cross}$. \label{fig:example1}} 
\end{figure*}

\begin{figure*}[t] 
   \centering
   \includegraphics[width=0.85\textwidth]{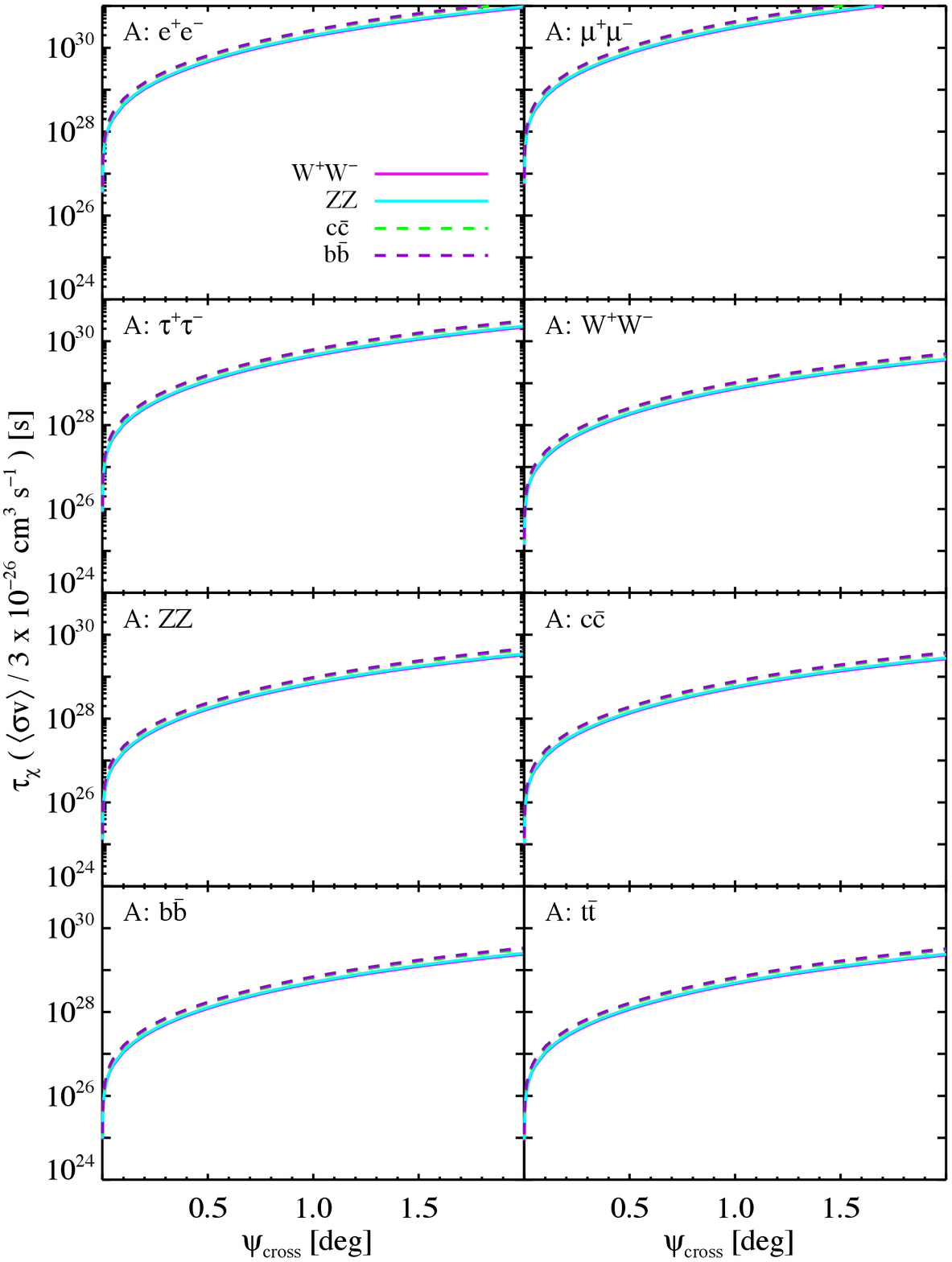} 
   \caption{Same as Fig.~\ref{fig:example1}, for hadronic and gauge
     boson decay channels (as labeled).  The four decay channels shown
     in each panel result in very similar curves due to their similar
     photon yields above 1~GeV for this particle
     mass.\label{fig:example2}}
\end{figure*}

Figs.~\ref{fig:example1} and~\ref{fig:example2} indicate the range of
dark matter parameters which would induce a transition between
annihilation and decay in the angular range of $0^{\circ}$-$2^{\circ}$
in Draco, (similar results are obtained for the other two dwarf
galaxies).  Here we neglect the contribution from substructure.  The
curves indicate the value of the dark matter lifetime at which the
intensities from dark matter annihilation and decay integrated above 
1~GeV are equal at $\psi_{\rm cross}$ (see Eq.~\ref{eq:taucross}, with
the integrated photon yield above 1~GeV).  The results for dark matter
decay into leptonic and semileptonic channels are shown in
Fig.~\ref{fig:example1}, and for decay into hadronic and gauge boson 
channels in Fig.~\ref{fig:example2}.  The annihilation channel for
each panel is labeled.  In these figures we assume a 200~GeV dark
matter candidate and an annihilation cross section in terms of
$\langle \sigma v \rangle$ typical of thermal dark matter, $\langle
\sigma v \rangle = 3 \times 10^{-26}$~cm$^3$~s$^{-1}$.  A larger cross
section would displace the curves downwards, scaling $\tau_\chi$
inversely to the cross section.  For a given $\psi_{\rm cross}$, above
the curves annihilation dominates and the emission profile is steeper,
while below the curves the dominant contributor is decay and the
profile is shallower.

The normalization of the curves depends on the relative photon yields
from annihilation and decay: for a given lifetime, the
annihilation-to-decay transition occurs further from the center of the
dwarf galaxy for channel combinations in which the ratio of the photon
yields from annihilation to decay is larger.  In each panel,
corresponding to a single annihilation channel, the variation in the
amplitude of the curves reflects the different photon yields for the
decay channels shown.  Decay via any of the hadronic or gauge boson
channels produces almost identical curves since the photon yields
above 1~GeV from these channels are similar, and these curves have the
highest normalization of any of the channels since their photon yields
are the highest. Similarly, there is little difference between the
curves for decay into any of the three semi-leptonic channels, and
these curves fall below the hadronic and gauge boson decay channel
curves.  The curves for decay into the leptonic channels show more
variation due to the larger variation in photon yields for these
channels, and as expected, fall below those for semi-leptonic and
hadronic and gauge boson channels due to their relatively low photon
yields.

For this energy threshold and an assumed cross section of $\langle
\sigma v \rangle = 3 \times 10^{-26}$~cm$^3$~s$^{-1}$, in order for
the transition to occur at an angle between $\sim 0.1^{\circ}$ and
$\sim 2^{\circ}$, the dark matter lifetime must be between $\sim
10^{25}$~s and $10^{31}$~s, depending on the combination of channels.
Let us note that each individual line in Figs.~\ref{fig:example1}
and~\ref{fig:example2} extends only over two orders of magnitude.
However, this alone does not tell about the sensitivity to the
lifetime and annihilation cross section as the range of parameters
depend on the particular combination of channels, which is not known. 
The range of values for the lifetime (for a given annihilation cross
section) we mention takes into account our ignorance in this
regard. For larger values of the annihilation cross section,
correspondingly smaller values of the lifetime are needed.

\begin{figure*}[t] 
   \centering
   \includegraphics[width=0.9\textwidth]{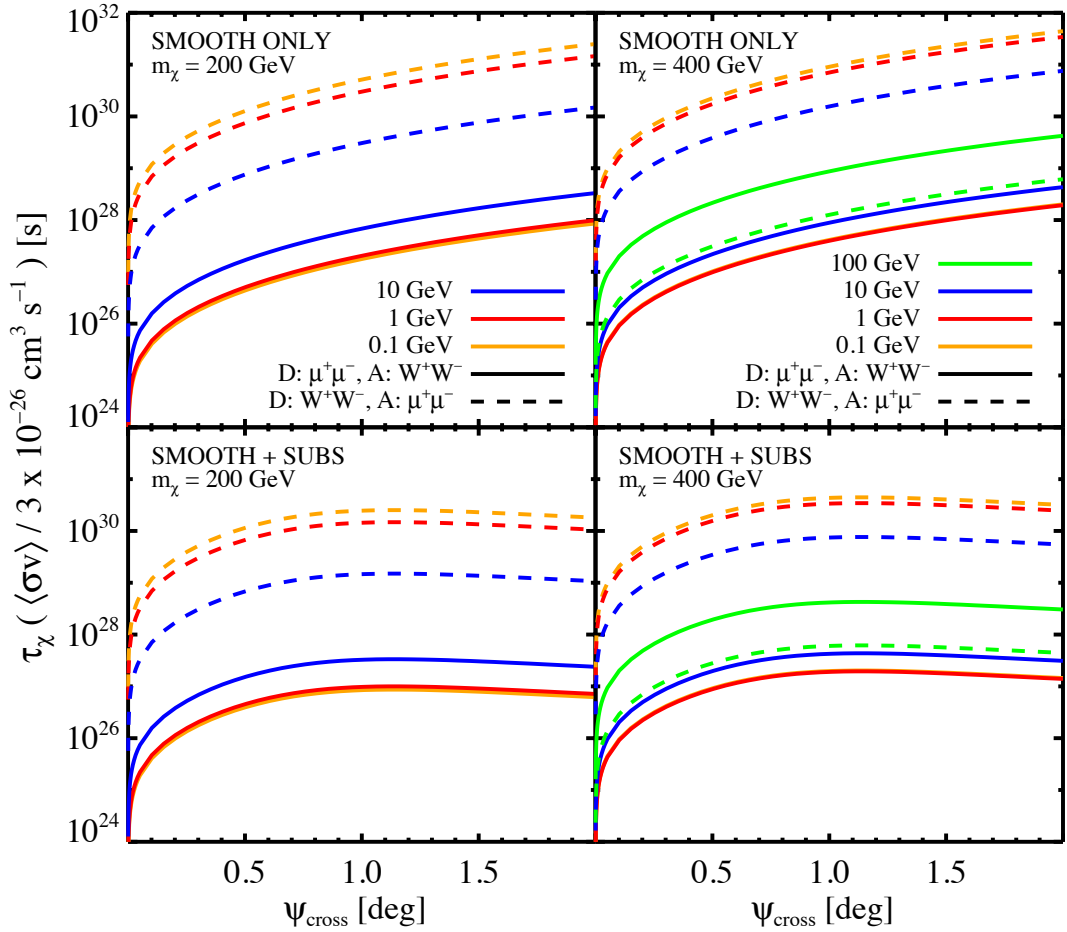}
   \caption{Effect of particle mass, energy threshold, and
     substructure on the lifetime $\tau_\chi$ at which the intensities
     from annihilation and decay are equal at $\psi_{\rm cross}$ from
     the center of Draco.  Each panel shows the results for a
     hard-soft ({\it solid curves}) and soft-hard ({\it dashed
       curves}) combination of decay-annihilation channels, with
     $\mu^{+}\mu^{-}$ representing a hard channel and $W^{+}W^{-}$ a
     soft channel.  In the left panels $m_\chi=200$~GeV while in the
     right panels $m_\chi=400$~GeV.  The top panels show the curves
     for the smooth halo only and the bottom panels include
     substructure.  The results for several choices of energy
     threshold are shown (as labeled); the curves corresponding to the
     100~GeV energy threshold are absent in the left panels since this
     is the largest possible energy of a photon from the decay of a
     200~GeV particle. 
 \label{fig:example3}} 
\end{figure*}

In Fig.~\ref{fig:example3}, we study the effect of adding substructure
and varying the energy threshold of the experiment for two different
dark matter masses, $m_\chi=$~200~GeV (left panels) and
$m_\chi=$~400~GeV (right panels).  We depict two combinations of 
channels: solid (dashed) lines represent the case of decay into a hard
(soft) channel and annihilation into a soft (hard) one.  As before, we
have chosen $\mu^+ \mu^-$ and $W^+ W^-$ as representative of hard and
soft channels, respectively.  In the upper panels we show the results
when only the smooth halo contribution is considered and in the lower
panels those when substructure is also present.  The different colors
represent several energy thresholds.  For $m_\chi =$~200~GeV, we have
not considered the threshold of 100~GeV, as this is the maximum energy
of the photons from the decay of a dark matter candidate of that mass.  

In this figure, the trend explained above regarding photon yields is
clearly evident: for the soft-hard combination of channels for
annihilation-decay, the transition in the angular emission profile
occurs further away from the center than in the hard-soft case.  We
see that for these masses, there is little difference between using an
energy threshold of 0.1~GeV or 1~GeV. However, a higher energy
threshold suppresses the contribution of the soft channel and thus the
curves are displaced upwards (downwards) if decay is into a hard
(soft) channel and annihilation into a soft (hard) one.  

In agreement with Fig.~\ref{fig:iprofiles}, Fig.~\ref{fig:example3}
shows that the effect of substructure starts to be important if
$\psi_{\rm cross} \gtrsim$~1$^{\circ}$.  As it is a very minor
correction to the signal from dark matter decay, substructure
contributes significantly only to the annihilation signal, and hence
it bends the curves in the lower panels downward, indicating that
annihilation remains the dominant source of the signal at larger
radii.  The presence of substructure results in the curves becoming
nearly horizontal for $\psi_{\rm cross} \gtrsim 1^{\circ}$, i.e., the
same ratio of annihilation to decay intensity is maintained at all
radii larger than this value.  This can be understood by recalling that 
the substructure contribution to the angular emission profile produces
a flattening at large angles very similar to that produced by dark
matter decay.  Hence, as a first approximation, for $\psi_{\rm cross}
\gtrsim$~1$^{\circ}$, there is a limiting value of the lifetime above
which decay is always subdominant in the intensity.  If the dark
matter lifetime is close to that value, the contribution from dark
matter decay and annihilation are comparable for $\psi_{\rm cross}
\gtrsim1^{\circ}$.

Since the required lifetime to produce a transition at a given angle
scales inversely with the annihilation cross section, these figures
indicate the ranges of the annihilation cross section and lifetime for
which the transition in the dominant component of the intensity from
annihilation to decay occurs within $\sim 0.1^{\circ}-2^{\circ}$ of
the center of the dwarf galaxy.  Interestingly, current constraints
from gamma-ray observations on the decay lifetime place a lower limit
of $\sim 10^{25}$ - $10^{26}$~s depending on the assumed mass and
channel~\cite{Chen:2009uq, Cirelli:2009dv, Papucci:2009gd,
  Zhang:2009ut, Essig:2009jx} (for earlier radiative bounds on the 
dark matter lifetime in different mass regimes see
Refs.~\cite{decayboundsrad} and for model-independent bounds see
Refs.~\cite{PalomaresRuiz:2007ry, decayboundscosmo}), a range which
overlaps slightly with the lower end of the range that would produce a
transition within $\sim 1^{\circ}$ of the center for the dwarfs
considered here for a thermal annihilation cross section.  Radiative 
bounds generally constrain the annihilation cross section to be
smaller than a factor of 10 - $10^4$ times a thermal cross section,
again depending on the mass and channel~\cite{Cirelli:2009dv,
  Papucci:2009gd, Abdo:2010ex, sigmaboundsrad} (see
Refs.~\cite{sigmaboundsnu, sigmaboundscosmo} for other bounds obtained  
using different techniques).  The parameters explored in the examples
presented here are therefore broadly representative of scenarios
allowed by current data and, moreover, include regions of parameter
space that may become accessible in the near future.

\section{Discussion and Conclusions}
\label{conclusions}

In this work we have outlined a strategy to constrain dark matter
properties in the event of the clear detection of an indirect signal 
from gamma-ray observations of dwarf galaxies.  We addressed the
question of how scenarios of dark matter annihilation, decay, or both
could be distinguished, and what information could be obtained about
the intrinsic properties of the dark matter particle and its
small-scale distribution from this type of indirect measurement.  

In principle, the indirect detection of dark matter in gamma-ray 
experiments would provide the energy spectrum of the signal.  The 
spectrum from dark matter annihilation has a maximum energy equal
to the dark matter particle mass, while the spectrum from decay has a
cutoff at half the mass.  Consequently, spectral information alone is
insufficient to identify the process that produced the signal.
However, in addition to the difference in the endpoint of the energy
spectrum, annihilation and decay give rise to different angular
distributions of the intensity of the emission.  Using this
information, we demonstrated that if a dark matter signal is clearly 
detected from a dwarf galaxy, an analysis of the energy spectrum of
the emission as a function of the angular distance from the center of
the object could provide the necessary information to distinguish the
cases of annihilation, decay, or both.   

The technique we propose, which combines spectral and angular
information, is particularly important due to the uncertainties in the
presence and properties of substructure in a dwarf galaxy halo.  In
particular, whereas the annihilation rate scales as the square of the
dark matter density, the decay rate depends linearly on the density.
However, the angular distribution of the annihilation (or decay)
signal from dark matter substructure also scales approximately
linearly with the smooth halo density in the outer regions of the
object, so its angular distribution roughly mimics that from dark
matter decay in the smooth halo.  Furthermore, since the amplitude of
the signal from annihilation in substructure depends sensitively on
the properties of the subhalo population and is not fixed by the
smooth halo density profile, the relative amplitude of the
annihilation signals from substructure and the smooth halo is
effectively independent (e.g., Ref.~\cite{Kuhlen:2008aw}).  Thus,
identifying annihilation in the smooth halo based on the emission
profile in the central regions of the object is not sufficient to 
distinguish between annihilation in substructure or decay as the
origin of emission from the outer regions of the halo.

In order to break this degeneracy, we have studied the energy spectrum
as a function of the angular distance from the center of the object
(see Figs.~\ref{fig:annuli} and~\ref{fig:inout}).  If a flattening in
the radial distribution of the intensity of the signal is observed at
an angular distance $\psi_{\rm cross}$ (c.f. Eq.~\ref{eq:taucross}),
this would point either to a significant contribution from both dark
matter annihilation and decay or to only dark matter annihilation with
an important contribution from substructure.  We have shown that a
change in the energy spectrum along with the change of slope in the
angular distribution could provide the necessary information to
confirm or reject the presence of a signal from dark matter decay.

If dark matter decay is established by an observed spectral change as
a function of angle, the signal in the innermost parts could be
studied to provide information about the annihilation contribution,
and that in the outermost regions (beyond $\psi_{\rm cross}$) would
provide information about the decay signal.  In principle, determining
$\psi_{\rm cross}$ and the intensity of the signal in the inner
(annihilation-dominated) and outer (decay-dominated) regions, could
help to constrain $\langle \sigma v \rangle$, $\tau_\chi$ and
$m_\chi$.  In this case, limits on the properties of substructure
could also be placed.

On the other hand, if beyond $\psi_{\rm cross}$ the energy spectrum
does not change, we expect annihilation in substructure to be the
cause of the flattening in the radial distribution and attribute the
signal to annihilation.  In this case, we could determine to some
extent the contribution of substructure and begin to constrain its
properties, such as the mass function slope, minimum subhalo mass, and
structural properties.  The indirect measurement would then yield the
particle mass and annihilation cross-section, and the absence of a
strong decay signal would enable further limits on decaying dark
matter to be placed.

Finally, if decay (annihilation) in the smooth halo can be established
as the dominant contribution due to a very shallow (steep) emission
profile at all angles, then the mass and lifetime (annihilation
cross-section) could be determined, and upper (lower) bounds placed on
the annihilation cross-section (lifetime).

In addition, as it is known and we have shown in \S\ref{spectra}, to a
first approximation the annihilation and decay spectra can be
classified as soft (due to hadronic or gauge boson channels) or hard 
(due to leptonic channels).  Hence, once the origin of the signal in
the different regions is established (annihilation in the smooth halo,
annihilation in the smooth halo and substructure, or decay), further
information can be obtained about the annihilation and decay channels
by studying the energy spectra.  On general grounds, in this situation
the distinction between dominantly leptonic or dominantly hadronic and
gauge boson channels could be achieved.

Our example scenarios focused on the case of a signal from Milky Way
dwarf spheroidal galaxies.  However we note that, in principle, this
method could also be applied to the case of our own galaxy.  In that
case an appropriate treatment of the secondary photons produced by the
prompt electrons and positrons via inverse Compton scattering off the
ambient photon background is necessary (see Ref.~\cite{Boehm:2010qt}
for a very recent related study).  In the case of our galaxy, this
contribution is very important (and it could be the dominant one) due
to the large light emission by stars and the infrared light as a
result of the scattering, absorption and re-emission of absorbed
starlight by dust. In addition, in order to recover any information
from an observation, the different backgrounds for a potential signal
would have to be properly addressed.  The galactic center is a complex
region, which makes the modeling of these backgrounds a difficult
task. All in all, bearing in mind these differences, a similar
application of the methodology described here could be possible in the
case of the Milky Way.

We have demonstrated the proposed method for different scenarios that
could also potentially explain the rise in the positron fraction
observed in the PAMELA data~\cite{PAMELA}
(Fig.~\ref{fig:inoutPAMELA}).  In the case that these dark matter 
models produce a detectable gamma-ray signal from dwarf galaxies, we
see that if the dark matter interpretation is correct, the observation
of the gamma-ray energy spectrum at different angular distances from
the center of the dwarf galaxies could help to establish if the origin
of the observed positron excess is due to dark matter annihilation,
decay, or both.

Finally, in Figs.~\ref{fig:example1},~\ref{fig:example2},
and~\ref{fig:example3} we have shown the range of dark matter
parameters for which a transition between annihilation and decay would 
occur within $2^\circ$ of the center of Draco (similar results are
obtained for Ursa Minor and Sagittarius).  We have presented the
results for different annihilation-decay combinations and see that the
dark matter lifetime must be in the range $\sim (10^{25}- 10^{31}$~s)
$(3 \times 10^{-26}$~cm$^3$~s$^{-1} / \langle \sigma v \rangle$), the
actual (narrower) range depending on the combination of annihilation
and decay channels.  The effect of substructure in the dwarf galaxy
halo is shown to be important only for $\psi_{\rm cross} \gtrsim
1^{\circ}$. The choice of energy threshold, however, strongly affects
the range of $\tau_\chi$ and $\langle \sigma v \rangle$ probed by this
technique (Fig.~\ref{fig:example3}), suggesting that, in the event of a
detection, observations covering a large energy range may be able to
explore a substantial region of the dark matter annihilation and decay
parameter space.

In order to apply the technique proposed in this study, the firm
detection of a gamma-ray signal of dark matter origin would be
required.  Although the main idea of this work is to show the
different potential features of a dark matter signal and to describe 
the method in a detector-independent way, we think it is worthwhile to
add a short discussion along these lines.  

The main point is to have access to two types of features, those in
the angular distribution and those in the energy spectrum.  In order
to reconstruct the energy spectrum in a given angular bin, a minimum
number of photons is necessary.  The integrated number of photons in
each angular bin scales with the solid angle $\Omega$ of the bin,
i.e., $\Omega \sim \pi (r_{\rm out}^2 - r_{\rm in}^2)$, where $r_{\rm
  out}$ and $r_{\rm in}$ are the angular radii of the outer and inner
edge of each bin.  For the equal width annuli used here, $\Omega$
increases linearly with the radius of the bin center, i.e., the
annulus centered on 0.85$^\circ$ encompasses 17 times as much solid
angle as the annulus centered on 0.05$^\circ$.  Let us assume, for
simplicity, that the difference in the total intensity between inner
and outer annuli is of the order of 10 to 100
(cf.~Fig.~\ref{fig:annuli}).  The photon flux from within the outer 
annulus is therefore typically a factor of a few smaller than from the
innermost annulus, if the spectral shapes in each annulus are similar.
Although the diffuse backgrounds would also scale with solid angle, in
the case that the backgrounds are small, sensitivity to a signal a
factor of only a few smaller than the signal in the innermost bin
would be required to measure the angular dependence of the integrated
flux.  If the backgrounds are large, sensitivity to signals a factor
of 10 to 100 smaller than that in the innermost bin would be necessary. 

While the spectra of the inner and outer annuli depend on the specific
combination of channels, very roughly we can say the energy spectrum
goes as $E^{-2}$ over most of the relevant energy range (see, e.g.,
Fig.~\ref{fig:annuli}), and hence the integrated photon yield over
some range scales approximately as $E^{-1}$.  The examples shown in
this study use the energy spectrum over 1 to 2 decades in energy, so
for equal logarithmic bins in energy, the flux at high energies is
$\sim$ 1 - 10\% of the flux at low energies.  To reconstruct the
energy spectrum, one could require, e.g., a few signal photons in the
highest energy bin of the outermost annulus.  This optimistically
implies that $\mathcal{O}(1000)$ total photons (angle- and
energy-integrated) within a 1 degree radius of the source are needed.
On general grounds, to get angular information a factor of a few more
statistics is necessary than are needed to detect the dark matter
signal clearly (i.e., with spectral info) in the first place.  Thus, if 
a signal in current experiments were detected just beyond the current
limits, in the future they might have the ability to do a study of
this kind, as the signal will improve with exposure.  While this
coarse estimate does not include a proper treatment of backgrounds,
which depend strongly on energy and on the experiment under
consideration, we note that the expected value of the photon counts
measured by \emph{Fermi} due to the Galactic diffuse gamma-ray
background above 1~GeV in a region of 1~degree around the center of
the dwarfs considered here after five years is $\sim$~100 - 1000.

In summary, we have shown that a dark matter particle with an
annihilation cross-section and lifetime just beyond the limits
currently established could produce a clear spectral change on an
angular scale within the reach of future experiments.  Ongoing
observations by \emph{Fermi}, HESS, VERITAS, and MAGIC and future
observations by the planned CTA and AGIS experiments will continue to
improve the prospects for detecting and mapping a dark matter signal
in the coming years.

\ack
We thank Joakim Edsj\"o and Torbj\"orn Sojstrand for help with PYTHIA
and Jorge Pe\~narrubia for useful communications. SPR thanks the
CCAPP, where this work started, and JSG thanks the CFTP for
hospitality. SPR is partially supported by the Portuguese FCT through
CERN/FP/83503/2008, CERN/FP/109305/2009 and CFTP-FCT UNIT 777, which
are partially funded through POCTI (FEDER), and by the Spanish Grant
FPA2008-02878 of the MICINN\@. JSG acknowledges partial support from
NSF CAREER Grant PHY-0547102 (to John Beacom).

\end{document}